\documentstyle[12pt,epsf,epsfig,rotate]{article}
\newcommand{\la}[1]{\label{#1}}
\newcommand{\be}{\begin{equation}}
\newcommand{\ee}{\end{equation}}
\newcommand{\ba}{\begin{eqnarray}}
\newcommand{\ea}{\end{eqnarray}}
\newcommand{\bi}{\begin{itemize}}
\newcommand{\ei}{\end{itemize}}
\newcommand{\rmi}[1]{{\mbox{\scriptsize #1}}}
\newcommand{\fig}{Fig.~}
\newcommand{\nr}[1]{(\ref{#1})}

\newcommand{\Hc}{{\rm H.c.\ }}
\newcommand{\nn}{\nonumber \\}
\newcommand{\fr}[2]{{\frac{#1}{#2}}}

\newcommand{\eq}{Eq.~}
\newcommand{\eqs}{Eqs.~}
\newcommand{\tb}{\tan\!\beta}
\newcommand{\href}[2]{#2}

\def\lsi{\raise0.3ex\hbox{$<$\kern-0.75em\raise-1.1ex\hbox{$\sim$}}}
\def\gsi{\raise0.3ex\hbox{$>$\kern-0.75em\raise-1.1ex\hbox{$\sim$}}}
\newcommand{\lsim}{\mathop{\lsi}}
\newcommand{\gsim}{\mathop{\gsi}}
  
\begin{document}
 
\begin{titlepage}
\begin{flushright}
CERN-TH/99-361\\
HD-THEP-99-22\\
hep-ph/9912278\\
\end{flushright}
\begin{centering}
\vfill
 
{\bf CP VIOLATING BUBBLE WALL PROFILES}

\vspace{0.8cm}

S.J. Huber$^{\rm a,}$\footnote{s.huber@thphys.uni-heidelberg.de}, 
P. John$^{\rm a,}$\footnote{p.john@thphys.uni-heidelberg.de}, 
M. Laine$^{\rm b,c,}$\footnote{mikko.laine@cern.ch} and
M.G. Schmidt$^{\rm a,}$\footnote{m.g.schmidt@thphys.uni-heidelberg.de} \\

\vspace{0.3cm}
{\em $^{\rm a}$Institut f\"ur Theoretische Physik, 
Philosophenweg 16, 
D-69120~Heidelberg, Germany\\}
\vspace{0.3cm}
{\em $^{\rm b}$Theory Division, CERN, CH-1211 Geneva 23,
Switzerland\\}
\vspace{0.3cm}
{\em $^{\rm c}$Department of Physics,
P.O.Box 9, FIN-00014 University of Helsinki, Finland}

\vspace{0.7cm}
{\bf Abstract}
 
\end{centering}
 
\vspace{0.3cm}\noindent
We solve the equations of motion for a CP violating phase between
the two Higgs doublets at the bubble wall of the MSSM electroweak 
phase transition. Contrary to earlier suggestions, we do not 
find indications of spontaneous ``transitional'' CP violation in the MSSM. 
On the other hand, in case there is explicit CP violation in the stop and 
chargino/neutralino sectors, the relative phase between the Higgses
does become space dependent, but only mildly even in the maximal case. 
We also demonstrate that spontaneous CP violation within 
the bubble wall could occur, e.g., if the Higgs sector 
of the MSSM were supplemented by a singlet. Finally we point 
out some implications for baryogenesis computations. 
\vfill
\noindent

\vspace*{1cm}
 
\noindent
CERN-TH/99-361\\
HD-THEP-99-22\\
December 1999

\vfill

\end{titlepage}
 
\paragraph{Introduction.}

For producing the baryon asymmetry of the Universe, the Sakharov
conditions demand three properties of a model. The first
one, baryon number non-conservation, is already present in the
Standard Model. The second one, deviation
from thermal equilibrium, can be potentially realized 
by a strongly first order electroweak phase transition. 
For the Higgs masses allowed 
this requires extensions of the Standard Model~\cite{isthere}, 
and the simplest such possibility appears to be the MSSM with a lightest 
stop lighter than the top~\cite{bjls}--\cite{mssmsim}\footnote{Among 
alternative scenarios leading to a strong 
transition is the MSSM augmented by a gauge singlet 
(NMSSM)~\cite{PietroniNMSSM}--\cite{HuberSchmidt}.}.
The third requirement, CP violation, comes into the game
when actually computing the baryon asymmetry. Many 
computations suggest that new CP violating phases as large as
${\cal O}(10^{-1})$ are required for generating the 
observed baryon asymmetry~\cite{hn}--\cite{risa}, and phases of this order 
of magnitude might potentially be in conflict with constraints coming 
from the electric dipole moment 
experiments~\cite{edm}\footnote{There is a recent
interest in scenarios where this conclusion can be avoided;
for a discussion and references see, e.g., \cite{pw}. We may note, 
in particular, that it appears sufficient to have the 1st and 2nd 
generation scalar partners heavy~\cite{ko}, an assumption 
often made in electroweak phase transition studies anyway.}.

Whether or not explicit phases are eventually a problem,  
it is in any case interesting to note that the MSSM may also offer 
a mechanism for generating enough CP violation for baryogenesis, 
without conflicting with {\em any} experimental constraints. Indeed, 
due to the fact that there are two Higgs doublets, one can 
in principle have a spontaneously generated
CP violating phase between them~\cite{lee}.
While spontaneous CP violation is 
excluded at $T=0$ for the experimentally allowed
parameter values~\cite{Pomarol}, 
there is a suggestion that
it might be more easily realized at finite temperatures~\cite{emq}, 
or even only in the phase boundary between the symmetric and broken
phases~\cite{cpr,fkot}. Such a profile could conceivably be quite 
useful for electroweak baryogenesis~\cite{mstv}. 

Let us stress that even if
explicit CP phases are present, 
it is important to know whether there
is some dynamics present in the system which 
intensifies or suppresses the explicit
effects around the phase transition. 
Thus we need to solve for the profiles
of the bubbles.

In order to really compute the profiles
and the baryon number produced 
at the electroweak phase transition, we should
follow the history of bubbles from the 
moment of nucleation, until the time the broken phase fills
the Universe. After nucleation, 
there is in general a long period of stationary
growth at a relativistic velocity, and then, 
if the latent heat of the transition is large enough to 
reheat the Universe back to the critical temperature $T_c$, 
another period of slower growth at a rate determined by the
expansion of the Universe~\cite{rth}. The stage of stationary
fast growth is characterized by a non-trivial hydrodynamical
temperature and velocity profile affecting 
also the Higgs field profiles~\cite{hydro}. We will not consider 
this problem here, but concentrate rather on the profile 
of a (nearly equilibrium) planar phase boundary at $T_c$
after the assumed reheating. 

Previously, the moduli of the two Higgs doublets around
the phase boundary at $T_c$ (and also for the newly 
nucleated bubble at $T<T_c$ before the setup
of a stationary hydrodynamical solution~\cite{mqs}) 
have been determined from the 2-loop effective 
potential~\cite{cm,mqs,pj}. The CP violating 
phase between the two Higgs doublets has been addressed in~\cite{fkot}. 
Both problems can in principle also be studied non-perturbatively 
with lattice simulations~\cite{cplr,latt}. 

The purpose of this paper is to present the 
first complete solution of the equations
of motion for the phase between the two Higgs doublets within 
the MSSM, utilizing a perturbative effective potential, 
but without restricting it to the effective quartic couplings. 
Our conclusions will differ from those obtained earlier on. 

\paragraph{Solving for the CP violating phase.}

We parameterize the two Higgs doublets of the MSSM as  
\be
H_1 = \frac{1}{\sqrt{2}} 
\left(
\begin{array}{l}
h_1 e^{i\theta_1} \\
0
\end{array}
\right), \quad
H_2 = 
\frac{1}{\sqrt{2}}
\left(
\begin{array}{l}
0 \\
h_2 e^{i\theta_2}
\end{array}
\right). \la{Hparam}
\ee
Since we want to use equations of motion involving only the Higgs
degrees of freedom, we have to make sure that no source
terms are generated for the gauge fields. The form of \eq\nr{Hparam}
guarantees that this is true for $W^\pm$ and the photon, and to remove
also the source terms for $Z$, we need to impose the constraint
\be
h_1^2 \partial_\mu \theta_1 = h_2^2\partial_\mu \theta_2. \la{constr}
\ee
In addition, because of gauge invariance, the effective Higgs 
potential depends on the phases only via $\theta=\theta_1+\theta_2$.  
We can then concentrate on $\theta$, and using \eq\nr{constr}
as well as assuming tree-level kinetic terms and moving to a frame 
where the bubble wall is static and planar, the action to be minimized is
\be
S \propto \int dz \Bigl[
\fr12 (\partial_z h_1)^2 + 
\fr12 (\partial_z h_2)^2 + 
\fr12\fr{h_1^2h_2^2}{h_1^2 +h_2^2}(\partial_z \theta)^2 + V_T(h_1,h_2,\theta)
\Bigr],\label{action}
\ee
where $V_T(h_1,h_2,\theta)$ is the finite temperature effective
potential for $h_1,h_2,\theta$.  In general, we are solving the
equations of motion for $h_1,h_2,\theta$ following from this
action. In the numerical solution 
we use the method outlined in~\cite{pj} which 
deals with the minimization of a functional of 
the squared equations of motion.

At the first stage, we consider the case with no explicit CP phases, 
and ask whether 
a particular solution without CP violation ($\theta=0,\pi$), 
is in fact a local minimum of the action or not. Clearly, it
is not if
\be
m_3^2(h_1,h_2)\equiv
\frac{1}{|h_1h_2|}\left.\frac{\partial^2V_T(h_1,h_2,\theta)}
{\partial\theta^2}\right|_{\theta=0}<0, \label{constraint}
\ee
where we have divided by $|h_1h_2|$, 
assuming that this is non-zero. 
\eq\nr{constraint} is to be evaluated along the path found 
by  solving the equations of motion for $h_1,h_2$.
We have chosen the convention that $h_1$ can have either sign,
allowing us to consider only $\theta=0$.
For the case of the most general quartic
two Higgs doublet potential, \eq\nr{constraint} agrees 
with the constraint first written down by Lee~\cite{lee}, 
on which most of the investigations of 
spontaneous CP violation are based~\cite{Pomarol}--\cite{fkot}. However, 
\eq\nr{constraint} is true more generally, independent of the 
form of the potential $V_T(h_1,h_2,\theta)$. 

Now, the tree-level potential of the theory is
\be
V_\rmi{tree}=  
\frac{1}{2}m_1^2 h_1^2 + \frac{1}{2}m_2^2 h_2^2 + 
m_{12}^2 h_1 h_2\cos\theta 
+\frac{1}{32}(g^2+{g^\prime}^2)( h_1^2- h_2^2)^2, \la{tree}
\ee
where $g,g'$ are the SU(2) and U(1) gauge couplings, and
at tree-level
\be
m_{12}^2 = -\frac{1}{2}m_A^2\sin2\beta. \la{m12}
\ee 
It follows that $m_3^2(h_1,h_2) = (1/2) m_A^2 \sin\!2\beta > 0$, 
so that the minimum of the potential in the 
$\theta$ direction is at $\theta=0$. Thus, in order
to get spontaneous CP violation one needs radiative 
corrections which can overcome the tree-level term. 

There are various mechanisms by which $m_3^2(h_1,h_2)$
might decrease. One potentially useful correction can be obtained 
at finite temperatures 
in the limit $m_U^2 \ll (2 \pi T)^2 \ll m_Q^2$, where $m_U^2,m_Q^2$ 
are squark mass parameters. Then~\cite{cplr},
\be
m_3^2(0,0) \to 
\fr12 m_A^2 \sin\!2\beta - 
\fr14  \frac{A_t \mu}{m_Q^2} h_t^2 T^2. \la{m3eff}
\ee  
Here $h_t\approx 1$ is the top Yukawa coupling, 
and $A_t,\mu$ are squark mixing parameters, 
assumed real for the moment.
The temperature correction is seen to reduce $m_3^2$
for $A_t\mu > 0$.
However, this alone does not improve the situation very much, since
$A_t\mu/m_Q^2$ is constrained by stability bounds 
to be $\ll 1$, particularly for small $m_U^2$ (see, e.g., \cite{clm}).

Another possibility is to radiatively generate quartic couplings 
which then effectively modify $m_3^2(h_1,h_2)$ 
at finite $h_1,h_2$~\cite{emq}--\cite{fkot}.
However, these effects are 1-loop suppressed in magnitude and 
are thus also 
typically relatively small. Moreover, one must take into account 
that $h_1,h_2$ are not free parameters but are determined by 
the couplings of the theory
and by the temperature: they should be solved for from the 
equations of motion. In particular, it can happen that tuning
$m_3^2(h_1,h_2)$ towards zero at the same time takes $h_1h_2$ to 
zero~\cite{cplr}: then the division carried out in \eq\nr{constraint}
is not defined and the effective potential itself, where
$h_1h_2$ multiplies $m_3^2(h_1,h_2)$, need not
become more favourable to spontanoues CP violation. 
In fact, taking the solution of the equations of motion into 
account and considering radiatively generated quartic couplings, 
it was argued in~\cite{cplr} that there is effectively a much 
stronger constraint for obtaining spontaneous CP violation in the MSSM: 
$m_A^2 + \# T^2 \lsim \lambda_5 (h_1^2+h_2^2)$ with $\lambda_5$
an effective quartic coupling of magnitude $\lsim 0.01$.
This constraint cannot be satisfied in practice. 

Nevertheless, these considerations are based on the approximation
to the effective potential where only the quadratic and quartic
operators are considered. At finite temperatures around the
electroweak phase transition, important contributions come from 
infrared sensitive non-analytic contributions which are not of 
this form, and can affect spontaneous CP violation~\cite{cplr}.
Thus, it is important to solve the equations of motion more 
generally for the full effective potential. 

In this work, we will consider the full finite temperature 
1-loop effective potential of the MSSM. It is known that 2-loop
corrections are very important in the MSSM in general~\cite{e},
and bring the perturbative results~\cite{bjls}--\cite{lo} rather close 
to the lattice results~\cite{mssmsim}, allowing for larger 
values of $h_1,h_2$ in the broken phase. Nevertheless, for the present 
problem we find that even 1-loop effects are in most cases very small, 
so we do not expect qualitative changes from the 2-loop effects. 
Eventually, the problem can be studied non-perturbatively with 
lattice simulations~\cite{cplr,latt}.


\paragraph{A scan for spontaneous transitional CP violation.}

The tree-level part of the effective potential $V_T(h_1,h_2,\theta)$
is in~\eq\nr{tree}. 
In the resummed 1-loop contribution to $V_T(h_1,h_2,\theta)$,
we include the same particle species as, e.g., in~\cite{fkot}:
gauge bosons, stops, charginos and neutralinos. This introduces dependences 
on the trilinear squark mixing parameters $A_t$ and $\mu$ 
as well as on the squark mass parameters $m_Q^2$, $m_U^2$, and the U(1), 
SU(2) gaugino parameters $M_1$ and $M_2$. We work in the 
$\overline{\mbox{DR}}$ scheme and the Landau gauge, choosing
as the renormalization scale $\bar{\mu}=246$ GeV. The parameters
$m_1^2, m_2^2$ of the $T=0$ potential are renormalized such that 
the minimum is at $(v_1,v_2)= (\cos\!\beta,\sin\!\beta)\times 246$~GeV.
The parameter $m_{12}^2$ can be expressed in terms of the
CP odd Higgs mass $m_A$ in the standard way, including 1-loop corrections.

We do not make a finite temperature expansion as some of the particles
can be heavy. Rather, the 1-loop temperature part of 
the effective potential, 
\begin{equation}
V_1(T\neq 0) = \frac{T^4}{2\pi^2} \sum_i n_i\int_0^\infty 
dx x^2\ln \left(1\mp e^{-\sqrt{x^2+z_i}}\right),
\end{equation}
where $z_i=m_i^2/T^2$ and $n_i$ counts the degrees of freedom
(negative for fermions),  
is evaluated using a spline interpolation 
between the high and low temperature regions. 

We now wish to see whether the constraint in \eq\nr{constraint}
can be satisfied at the bubble wall between the symmetric
and broken phases. To do so, we have to search for each parameter
set for the critical temperature $T_c$, solve the equations
of motion for $(h_1,h_2)$ between the minima, and evaluate
$m_3^2(h_1,h_2)$ along this path. Since this is 
quite time-consuming, we proceed in two steps.

{\bf 1.} 
At the first stage, we do not solve for $h_1,h_2,T_c$, but rather
take them as free parameters in the ranges 
$h_1/T=-2..2$ and $h_2/T=0..2$, $T=80...120$ GeV.
The zero temperature parameters are varied in the wide ranges
\ba
& & \tan\beta  = 2...20, \qquad 
m_A = 0...400 \mbox{ GeV}, \nn
& & m_U =  -50...800 \mbox{ GeV}, \qquad
m_Q = 50...800 \mbox{ GeV}, \\
& & \mu, A_t, M_1, M_2 = -800...800  \mbox{ GeV}. \nonumber
\ea
Here a negative $m_U$ means in fact a negative 
right-handed stop mass parameter, $-|m_U^2|$. We have also
studied separately the (dangerous~\cite{cms}) region where 
the transition is very strong~\cite{bjls}--\cite{mssmsim}, 
corresponding to $m_U \sim -70...-50 \mbox{ GeV}$. 

Note that since we
do not solve for the equations of motion at this stage but 
allow for $h_1=\pm |h_1|$, we have to divide
in \eq\nr{constraint} by $h_1h_2$ instead of $|h_1h_2|$:
this leads in general to positive values due to the 
tree-level form of the potential, \eqs\nr{tree},\nr{m12}. 
A signal of a potentially promising region is then 
a small absolute value of the result, since this means 
that we are close to a point where $\partial_\theta^2 V_T(h_1,h_2,\theta)$
crosses zero. 

{\bf 2.}
At the second stage, we study the most favourable
parameter region thus
found in more detail. First of all, we search for the critical
temperature. Then, we solve the equations of motion for $(h_1,h_2)$. 
By comparing with the exact numerical solution in several cases, 
we find that a sufficient accuracy can be obtained in practice
by searching for the ``ridge'' as an approximation to the wall
profile. It is determined as the line of maxima 
of the potential in the direction
perpendicular to the straight line between the minima. 
Finally, we look for the minimum of $V_T(h_1,h_2,\theta)$
at fixed $(h_1,h_2)$: this is a fast and reliable approximation
for the full solution in the case that $\theta$ is small
(i.e., just starts to deviate from zero), and 
corresponds to \eq\nr{constraint}.

For the first stage, we perform a Monte Carlo scan  
with about $10^9$ configurations. 
Small values of $m_3^2(h_1,h_2)$
are scarce, and even then do not necessarily correspond to the desired
phenomenon of spontaneous CP violation: they could also be 
points far from the actual wall. This will be clarified at stage 2. 

\begin{figure}[tb]

\vspace*{-0.5cm} 

\centerline{\epsfysize=8cm\epsffile{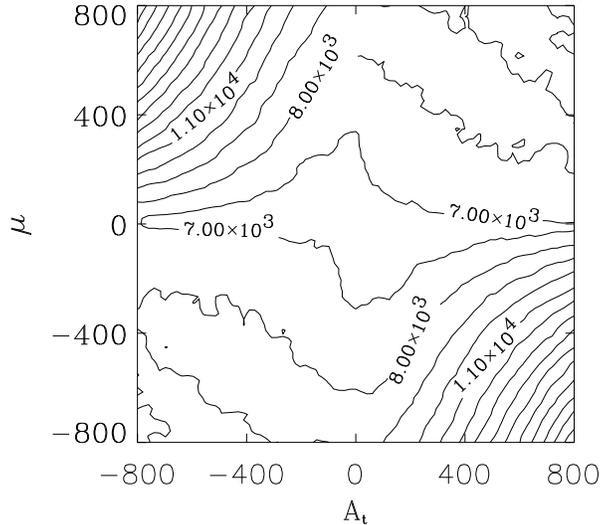}}

\vspace*{-0.5cm} 

\caption{The average value of $m_3^2$ versus $\mu$ and $A_t$. 
We observe that small values of $m_3^2$ are not typical in any part
of the plane but are on the average more likely for small $\mu,A_t$,
and that the distribution is wider (and thus more favourable) for
like signs of $\mu,A_t$, as shown by the noisy contours obtained
with a finite amount of statistics.}
\label{Atmu}
\end{figure}

The parameter region found depends most strongly on $m_A$, $\tb$, 
with a preference on small values of $m_A$ and large of $\tb$, 
such that $m_{12}^2$ in \eq\nr{m12} is small. (It can be 
noted that this requirement is not favourable for a 
strong phase transition~\cite{bjls}--\cite{lo}). There is also 
a relatively strong dependence on $A_t$ and $\mu$: the 
region favoured is shown in \fig\ref{Atmu}. The dependences on 
the other parameters are less significant; for
$m_U$ and $m_Q$ small values are preferred. The region found
is in rough agreement with those found in~\cite{emq,cpr,fkot,cplr}.

At the second stage, we make further restrictions.  
For instance, we
exclude the cases leading to non-physical negative mass 
parameters, e.g. a lightest stop $m^2_{\tilde t}<0$ (a stronger
restriction could be obtained by excluding regions leading to 
a charge and colour breaking minimum at some stage of the 
Universe expansion~\cite{cms}). We exclude cases leading to $T=0$ 
spontaneous CP violation in the broken phase: this 
phenomenon requires very small values of $m_A$ \cite{Pomarol}.
We also discard phase transitions which are exceedingly weak, 
$v/T\ll 0.1$. 

In any case, even before taking into account the experimental
lower limits on the Higgs masses, 
$m_H, m_A \gsim 80 $ GeV, we cannot find any 
promising case  in the sample of $\sim 2\times 10^6$ 
configurations of stage 2, 
with the desired property of a temperature induced transitional 
CP violation within the bubble wall in the MSSM. 

In~\cite{fkot}, the special point $m_U^2\approx 0$ was considered.
Due to the fact that in~\cite{fkot} thermal mass corrections were
neglected for $m_U^2$, this point corresponds in the physical MSSM
to the case where the thermally corrected stop mass parameter
vanishes, $m_U^2 + \#T^2\sim 0$. 
Expanding the 1-loop cubic term from the stops
to a finite order in $v_1/v_2$, it was suggested that transitional
spontaneous CP violation can take place. This region is quite 
dangerous due to the vicinity of a charge and colour breaking
minimum~\cite{cms}, and furthermore, perturbation theory is not
reliable. In any case, without expanding the 1-loop 
contribution in $v_1/v_2$, we cannot reproduce the behaviour
proposed in \cite{fkot}. 

We conclude that after taking into account the infrared sensitive 
effects inherent in the 1-loop effective potential, coming
from a light stop and gauge bosons, and solving for the wall
profile from the equations of motion, spontaneous CP violation
does not take place in the physical MSSM bubble wall. 

\paragraph{Explicit CP violation.}

We next turn to explicit CP violation. For the moment we ignore
the experimental constraints on the magnitude of the explicit CP phases.
Assuming universality in the gaugino sector, 
there are in principle four independent parameters which could
carry a phase: $m_{12}^2,A_t,\mu,M_2$. 
However, only two of the phases are physical after field
redefinitions: $\theta_A=\theta_{m_{12}^2}+\theta_{A_t}+\theta_\mu$ 
appearing in the stop matrix, and  
$\theta_C=\theta_{m_{12}^2}+\theta_\mu+\theta_{M_2}$ 
appearing in the chargino and neutralino matrices. In addition there  
is the dynamical phase between the two doublets, $\theta$.  
The mass eigenvalues entering the 1-loop effective potential 
now have to be computed in the presence of these complex phases. 

\begin{figure}[tb]

\vspace*{.cm}

\centerline{\epsfysize=8.5cm\rotate[r]{\epsffile{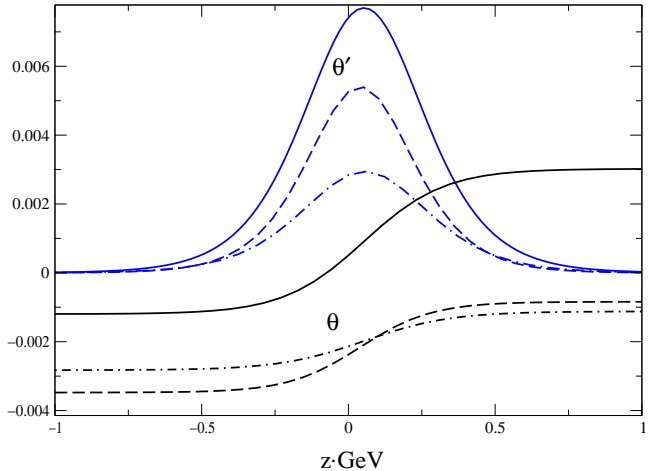}}}

\vspace*{.cm} 

\caption{The phase $\theta$ and its derivative $\theta'$ 
in the case of large explicit phases (see the text) for
three sets of $m_A,\tb$:  
$m_A=80$ GeV, $\tb=2.0$ (solid);
$m_A=120$ GeV, $\tb=2.0$ (dashed); 
$m_A=120$ GeV, $\tb=3.0$ (dot-dashed). We have
$m_U =0$ GeV, $T_c\approx 100$ GeV.}
\label{explicit2}
\end{figure}

We can again search for the minima along the $\theta$ direction,
as we have verified that 
this is a good approximation to the full solution
for the small values of $\theta$ we shall find. Indeed, even 
for maximal phases $\theta_A=\pi/2$ and
$\theta_C=\pi/2$, we find a strongly suppressed CP phase
$\theta$ in the broken Higgs phase.
For $m_A\gsim 80$ GeV, $\theta(x)$ is of order
$10^{-2}...10^{-3}$, and varies relatively mildly 
within the wall (Fig.~{\ref{explicit2})\footnote{As a technical point, 
it should be noted that the  determination of $\theta(x)$ 
is more difficult (and less meaningful) closer to the symmetric phase. 
In Fig.~\ref{explicit2}, the solution has
been obtained by using a $\tanh$-ansatz for $\theta(x)$, which turns out to 
compare very well with the more precise solution in the middle
of the wall, where the full equations can be more easily solved.}. 
Only for experimentally excluded small values of $m_A$ do we obtain 
phases up to order unity. 
For explicit phases of order ${\cal O}(10^{-1})$, 
the dynamical phase $\theta$ generated is typically very small, 
${\cal O}(10^{-3}-10^{-4})$, and thus, from the baryogenesis point 
of view, has an effect inferior to those arising from the explicit 
phases, ${\cal O}(10^{-1})$~\cite{non-eq}--\cite{risa}.

\paragraph{Transitional CP violation in the NMSSM.}

Finally, we will consider the case of the singlet extension 
of the MSSM, called the NMSSM. In this case there is a large
parameter region allowing for a strong phase 
transition~\cite{PietroniNMSSM}--\cite{HuberSchmidt}. 

The most general superpotential containing the two Higgs doublets
which are already present in the MSSM, and the gauge singlet field
$S$, can be written as~\cite{HaberKane} 
\begin{equation} \label{n1}  
W=\mu H_1H_2+\lambda SH_1H_2 -\frac{k}{3}S^3-rS.
\end{equation} 
In addition, there are soft SUSY breaking terms. As a result,
the mass parameter $m_{12}^2$ of the MSSM gets effectively
replaced by a dynamical variable, 
\be
m_{12}^2 H_1H_2 + \Hc \to
(m_{12}^2+ \lambda A_{\lambda}S +  \lambda k^* S^{2*}) H_1H_2 + \Hc. 
\ee 
Here the singlet $S$ is a complex field. It is thus clear that 
at the phase boundary where the singlet field can have a non-trivial
profile, the spontaneous phase $\theta$ between the two Higgs 
doublets will also have one (for an analysis of spontaneous
CP violation in the NMSSM, based on effective couplings, 
see~\cite{nmssm}). 

We show an example of the behaviour of the system, for a specific
choice of (real) parameters, in \fig\ref{wand_mit_5}. The singlet
has been written as $S=n+ic$.
All five fields, $n,c,h_1,h_2,\theta$, have been solved from the 
equations of motion, where the effective potential contains
1-loop contributions from the tops, stops, gauge bosons, charginos,
neutralinos and Higgs bosons \cite{HuberSchmidt,HuberProc}.
In the symmetric phase, the singlet carries a complex vev, while
in the broken phase it is real; correspondingly, the phase
$\theta$ is non-zero close to the symmetric phase but goes to 
zero in the broken phase. This demonstrates the general 
possibility of transitional CP violation in bubble walls 
of the NMSSM phase transition. 

\begin{figure}[tb]

\vspace*{-1.cm}

\centerline{\epsfysize=10.5cm\rotate[r]{\epsffile{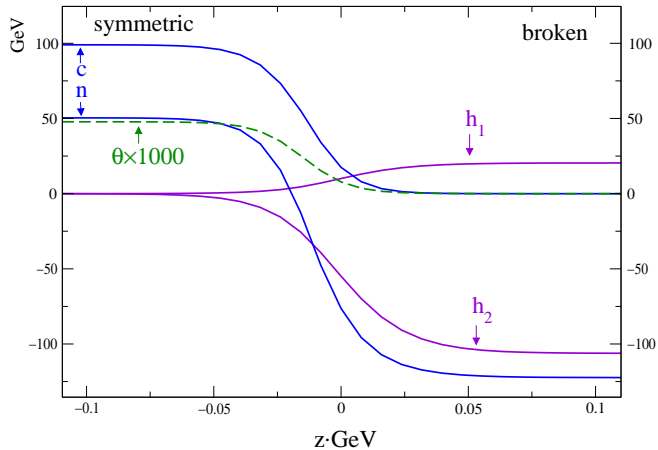}}}

\vspace*{-.5cm} 

\caption{Spontaneous
transitional CP violation in the bubble wall. 
A full solution with 5 fields 
is shown. The phase $\theta$ 
varies from a finite value in the symmetric phase to zero in the
broken phase.}
\label{wand_mit_5}
\end{figure}

\paragraph{Conclusions.}

The aim of this paper was the discussion of CP violation in the
MSSM bubble walls. We searched
in a rather large parameter space for transitional CP violation, 
improving on earlier determinations~\cite{cpr,fkot,cplr} by
using the full infrared sensitive
1-loop effective potential and solving the 
equations of motion. We
could not find any parameter set permitting temperature induced
transitional CP violation in the MSSM. 
The most favourable region (small $m_A$, large $\tb$) is also  
in contradiction with a strong first order
phase transition, as well as with experimental 
constraints. The dependences are dominated by the tree-level
parameters $m_A$ and $\tan\beta$. 

We have omitted the 2-loop corrections in the effective potential, 
as well as the effects from the Higgses, since we expect them to 
be small for the present problem dominated by tree-level effects. 
Eventually, this expectation can be checked with lattice 
simulations~\cite{cplr,latt}.

We also investigated the profile of the dynamical phase in 
the presence of explicit CP violation in the soft supersymmetry
breaking parameters of the MSSM. 
We found that the dynamical effects are 1-loop suppressed. 
Thus, even large explicit phases result in a
relatively small dynamical phase in the actual bubble 
wall at $T_c$, and it seems that the dynamical effects
are subdominant for baryogenesis.

Finally we showed an example of an 
NMSSM bubble wall with five fields
including two CP violating phases. This example demonstrates
transitional CP violation in a bubble wall with a CP conserving 
broken phase.

\paragraph*{Acknowledgements.}

M.L. thanks J.~Orloff for useful discussions. 
This work was partly supported by the TMR network 
{\em Finite Temperature Phase Transitions in Particle
Physics}, EU contract no.\ FMRX-CT97-0122.


\end{document}